# Methods for the inclusion of real-world evidence in network meta-analysis

David A Jenkins[1,2,3,*], Humaira Hussein[1,*], Reynaldo Martina[1], Pascale Dequen-O'Byrne[1], Keith R Abrams[1,4], Sylwia Bujkiewicz[1] on behalf of IMI GetReal Work Package 1

[1] Biostatistics Research Group, Department of Health Sciences, University of Leicester,

University Road, Leicester, LE1 7RH, UK

[2] School of Health Sciences, Faculty of Biology, Medicine and Health, University of Manchester, Oxford Road, Manchester, M13 9PL, UK

[3] NIHR Greater Manchester Patient Safety Translational Research Centre, University of Manchester, Oxford Road, Manchester, M13 9PL, UK

[4] Centre for Health Economics, University of York, York, YO10 5DD, UK

*David Jenkins and Humaira Hussein equally contributed to this work

## Abstract

**Background:** Network Meta-Analysis (NMA) is a key component of submissions to reimbursement agencies world-wide, especially when there is limited direct head-to-head evidence for multiple technologies from randomised controlled trials (RCTs). Many NMAs include only data from RCTs. However, real-world evidence (RWE) is also becoming widely recognised as a valuable source of clinical data. This study aims to investigate methods for the inclusion of RWE in NMA and its impact on the level of uncertainty around the effectiveness estimates, with particular interest in effectiveness of fingolimod.

**Methods:** A range of methods for inclusion of RWE in evidence synthesis were investigated by applying them to an illustrative example in relapsing remitting multiple sclerosis (RRMS). A literature search to identify RCTs and RWE evaluating treatments in RRMS was conducted. To assess the impact of inclusion of RWE on the effectiveness estimates, Bayesian hierarchical and power prior models were applied. The effect of the inclusion of RWE was investigated by varying the degree of down weighting of this part of evidence by the use of a power prior.

**Results:** Whilst the inclusion of the RWE led to an increase in the level of uncertainty surrounding effect estimates in this example, this depended on the method of inclusion adopted for the RWE. Power prior NMA model resulted in stable effect estimates for fingolimod *yet increasing the width of the credible intervals* with increasing weight given to RWE data. The hierarchical NMA models were effective in allowing for heterogeneity between study designs, however, this also increased the level of uncertainty.

**Conclusion:** The power prior method for the inclusion of RWE in NMAs indicates that the degree to which RWE is taken into account and can have a significant impact on the overall level of uncertainty. The hierarchical modelling approach further allowed for accommodating differences between study types. Consequently, further work investigating both empirical evidence for biases associated with individual RWE studies and methods of elicitation from experts on the extent of such biases is warranted.

## Background

When evaluating new health technologies, traditionally data from randomised controlled trials (RCTs) have been considered a gold standard and, as such, used in meta-analysis in the evaluation process of new health technologies. Recently, there has been a growing interest in the use of real-world evidence (RWE) from observational studies in health-care evaluation (Gami et al., 2007, Salpeter et al., 2009). This is particularly the case in rare disease areas or in conditions where RCT design may be less feasible. The inclusion of RWE in a network meta-analysis (NMA) of data from RCTs is not a straightforward issue, as the effectiveness estimates obtained from RWE may be subject to selection bias, due to lack of randomisation, and hence use of randomised evidence may be preferable. However, the potential advantage of RWE, particularly for the purpose of health technology assessment (HTA) decision-making, is that it can be a substantial source of evidence thus increasing the available evidence base as well as better representing "real-life" clinical practice. To this extent, RWE can be used to bridge a gap between efficacy and effectiveness to ensure that the evaluation process reflects what is expected in clinical practice in terms of effectiveness of new health technologies. Therefore, recent methodological developments focus on appropriate methods of using such data.

As part of the IMI GetReal initiative, aiming to incorporate real life clinical data into drug development, methodologies were investigated for including such data in the later stages of the drug development process (i.e. health technology assessment), where data on treatment effectiveness can be included in the meta-analysis to inform HTA decision-making (Makady et al., 2017).

A number of methods have been used to combine evidence from different sources, which include naïve pooling (Li and Begg, 1994), inclusion of external sources of evidence as prior information (Mak et al., 2009, Spiegelhalter et al., 2002), power transform prior approach (Ibrahim and Chen, 2000) and hierarchical modelling (Prevost et al., 2000). These methods were originally introduced in standard pairwise meta-analysis and later generalised by Schmitz et al. (2013) to network meta-analysis (NMA) to combine direct and indirect evidence from a number of studies investigating effectiveness of a number of treatments (Schmitz et al., 2013). NMA has been used routinely in technology assessments conducted by many HTA agencies world-wide. It is a particularly useful meta-analytic tool when data from head-to-head trials on an intervention of interest are limited. NMA is used to combine evidence from studies of heterogeneous treatment contrasts and is also known as mixed treatment comparisons meta-analysis.

The aim of this paper is to investigate the use of NMA to combine estimates obtained from both RCTs and RWE using methods that differentiate between the study designs to account for the potential inherent biases present in RWE. A range of methods for combining RCT data with RWE in an NMA setting are discussed, which include naïve pooling, hierarchical modelling and power transform prior approach. The hierarchical model has been extended here to include power transform priors. The methodology is applied to an illustrative example in relapsing-remitting multiple sclerosis (RRMS) (Canadian Agency for Drugs Technologies in Health, 2013). A systematic literature review was carried out to identify sources of data, from both RCTs and RWE, on the effectiveness of disease modifying therapies (DMTs) used in RRMS patients. The results from the review and extracted data were subsequently used to illustrate how the three methodologies can be used to combine the data from the two types of sources of evidence and to compare their impact on the treatment effect estimates and resulting uncertainty.

## Methods

### Illustrative example and sources of evidence

In a motivating example, DMTs used in patients with RRMS were considered. A systematic review was carried out to identify studies, both randomised and observational, of different DMTs with a main focus on effectiveness of fingolimod to illustrate how the inclusion of RWE in NMA would impact the estimates of effectiveness of fingolimod in the context of a technology appraisal. The literature search was limited to studies reported prior to January 2010, when fingolimod was given licencing authorisation. Data were extracted on the effect of each treatment on relapse rate. Search terms utilised is available in Additional File 1.

### Network meta-analysis

Random-effects NMA model with adjustment for multi-arm trials (Ades et al., 2007) was used as the base case meta-analytic model. To investigate the effect of fingolimod on relapse rate, the number of relapses $r_{ik}$ in each study *i* and treatment arm *k* was modelled as count data following the Poisson distribution (Crowther et al., 2012),

$$r_{ik} \sim Poisson(\gamma_{ik} E_{ik}) \quad [1]$$

where $E_{ik}$ is the exposure time in person years and $\gamma_{ik}$ is the rate at which events (relapses) occur in arm $k$ for study $i$. Following a standard generalized linear model approach, the conjugate log link was used with random true treatment effect differences $\delta_{ibk}$ between treatments $k$ and $b$ which are assumed to follow common normal distribution:

$$\log(\gamma_{ik}) = \mu_{ib} + \delta_{ibk} I_{k \neq b} \quad [2]$$

$$\delta_{ibk} \sim N(d_{bk}, \sigma^2) \quad [3]$$

Assuming consistency in the network (which means that, for example, average treatment effect difference $d_{AC}$, between treatments $A$ and $C$, equals the sum of average treatment effect differences $d_{AB,}$between treatments $A$ and $B$, and $d_{BC,}$between treatments $B$ and $C$) allows to represent treatment effect for each treatment contrast $d_{bk}$ in the network as a difference of basic parameters which are average treatment effects of each treatment in the network compared to a common reference treatment 1; $d_{bk} = d_{1k} - d_{1b}$. Adopting a Bayesian approach to estimating the parameters of equations [1] – [3] requires that prior distributions are placed on the model parameters: the baseline study effects, $\mu_{ib}$, for example, the uniform distribution $\mu_{ib} \sim Uniform(-10, 10)$, on the basic parameters, $d_{1k} \sim Uniform(-10, 10)$ and on the between-study variance $\sigma \sim Uniform(0, 2)$.

For multi-arm studies, correlation between treatment effects relative to a common baseline treatment is taken into account by assuming true treatment effects $\delta_{i(bk_n)}$ follow common multivariate normal distribution which can be represented as series of univariate conditional distributions as follows:

$$\delta_{i(bk_1)} \sim \text{Normal}\left(d_{(bk_1)}\,,\ \sigma^2\right) \qquad [4]$$

$$\delta_{i(bk_n)} | \begin{pmatrix} \delta_{i(bk_1)} \\ \vdots \\ \delta_{i(bk_{n-1})} \end{pmatrix} \sim \text{Normal}\left( d_{(bk_n)} + \frac{1}{n}\sum_{t=1}^{n-1}\left(\delta_{i(bk_t)} - d_{(bk_t)}\right), \frac{(n+1)}{2n}\sigma^2 \right) \qquad [5]$$

where $n = 2, \ldots, p$ in the $(p+1)$-arm study of $p$ treatment effect estimates relative to the reference treatment.

## Naïve pooling approach

The above NMA model was initially used to combine data from RCTs with RWE by including the observational studies at 'face-value'. Data from all studies, regardless of the study design, were combined in the NMA described above.

This model was then extended to account for the differences between the designs of the studies as described in the following sections.

## Power prior approach

To take into account the differences in  study design between RCTs and observational studies, a 'power transform prior' approach was adopted (Ibrahim and Chen, 2000). This approach allows to down-weight the RWE, thus making the data from this type of studies contribute less compared to data obtained from the RCTs. This is achieved by introducing a down-weighting factor, alpha (α), which the likelihood contribution of the RWE studies is raised to the power of.  Alpha (α) is then varied between zero and one, with zero meaning that RWE is entirely discounted in the NMA, and with one indicating that all RWE is considered at 'face-value', which is assumed to be the same for each RWE study included in the network, The impact of different levels of weighting on the results of the NMA is performed by considering a series of values for alpha. The results are then summarised both in terms of the effect estimates (and their associated level of uncertainty) and the rankings that the treatments received (based on these effect estimates).

Considering the annualised relapse rate ratio (ARRR) and assuming $\delta = \log(ARRR)$, the overall joint posterior distribution is given by,

$$P(\delta|RCT, RWE) \propto L(\delta|RCT) \times L(\delta|RWE)^{\alpha} P(\delta) \qquad [6]$$

where $L(\theta|Y)$ is the likelihood of $\theta$ given data $Y$. Assuming a standard random-effects NMA model, we combine the likelihood contribution of RWE, raised to the power of alpha, with the likelihood of the RCT data. Together with the prior distributions for the basic parameters, this gives the overall posterior distribution with RWE discounted by the parameter alpha. Assuming that the number of relapses follow a Poisson distribution, the RWE log likelihood (LL) in [1] becomes

$$LL_{ikh} = log\left(\frac{{\gamma_{ikh}}^{r_{ikh}} e^{-\gamma_{ikh}}}{r_{ikh}!}\right)^{\alpha_h} \qquad [7]$$

$$LL_{ikh} = \alpha_h(r_{ikh}\log(\gamma_{ikh}) - \gamma_{ikh} - \log(r_{ikh}!)) \qquad [8]$$

where $h$ indexes the different values of $\alpha$.

## Hierarchical model approach

An alternative approach to allowing differentiation between study designs in NMA is introducing another level in a Bayesian hierarchical model, modelling the between-study heterogeneity of treatment effects within each study design (RCT or RWE) and across study designs. Adapting the models introduced by Schmitz et al. (2013), the hierarchical model was adjusted to account for a Poisson distribution to assess the count data. Assuming $j = 1, 2$ where 1 represents the RCT data and 2 represents the RWE then equation [1] now becomes,

$$r_{ik}^{j} \sim Poisson(\gamma_{ik}^{j} E_{ik}^{j}) \qquad [9]$$

And, similarly as in the general NMA model, using the log link function equation [2] becomes

$$\log(\gamma_{ik}^{j}) = \mu_{ik}^{j} + \delta_{ibk}^{j} I_{k \neq b} \qquad [10]$$

The data from the two sources of evidence, RCT data and RWE data, are modelled separately at the within-study and within-design level. Similarly, as in Schmitz et al. (2013) assuming the RCT and RWE evidence are exchangeable, the study design estimates are combined to an overall measure of treatment effect using random-effects (Schmitz et al., 2013). Thus, if $\delta_{ibk}^{1}$ and $\delta_{ibk}^{2}$ represent the treatment effect of treatment $k$ against a reference treatment $b$, based on the RCT evidence and RWE respectively, then,

$$\delta_{ibk}^{1} \sim N(d_{bk}, \sigma^2) \qquad [11]$$

$$\delta_{ibk}^{2} \sim N(d_{bk}, \sigma^2) \qquad [12]$$

where $d_{bk}$ is the mean treatment effect of treatment $k$ compared to a reference treatment $b$ and $\sigma^2$ is the variance representing the design level between-study heterogeneity. Prior distributions need to be placed on the parameters of the model, for example, the following "vague" prior distributions,

$$d_{bk} \sim Uniform(-10,10)$$

$$\sigma \sim Uniform(0,2)$$

This model was further extended by adopting a power prior approach at the within-study level for RWE (level one) by down weighting the likelihood contribution of the RWE by the factor alpha, as in formula [6], in the hierarchal model in order to provide a further sensitivity analysis. Combining average underlying study effects $\delta_{ibk}^{1}$ from RCTs with down-weighted effects $\delta_{ibk}^{2}\ from\ RWE$ produces overall pooled ARRR combined effects $d_{bk}$.

## Implementation and model fit

All models were implemented in WinBUGS version 1.4.3 (Lunn et al., 2000). The first 10,000 simulations were discarded for all models as a burn-in. The main analyses were based on additional 20,000 iterations in order to ensure convergence. Convergence was investigated by visually inspecting the trace and history plots. Model fit was evaluated using the total residual deviance and the DIC for each network size (Spiegelhalter et al., 2002). Between-study heterogeneity was assessed using the standard deviation across random-effects models. Inconsistency was assessed by assessing residual deviance and performing node splitting analysis (Dias et al., 2010).

## Results

### Network structure

Figure 1 illustrates the network diagram of direct comparisons between interventions in both the RWE and RCT data. The nodes represent individual interventions analysed and the interconnecting lines represent the direct comparisons between interventions. The numbers along these lines represent the number of studies for each comparison in either the RCTs or RWE. In total there were 23 studies included, 14 of them being RCTs. One may expect the RWE studies to have a larger sample size. However, in our example the average sample size in each arm for the RWE was 186 participants, compared to the 288 participants in the RCT arms. The list of studies in the NMA is included in Additional file 2 with data extracted reported in Additional file 3.

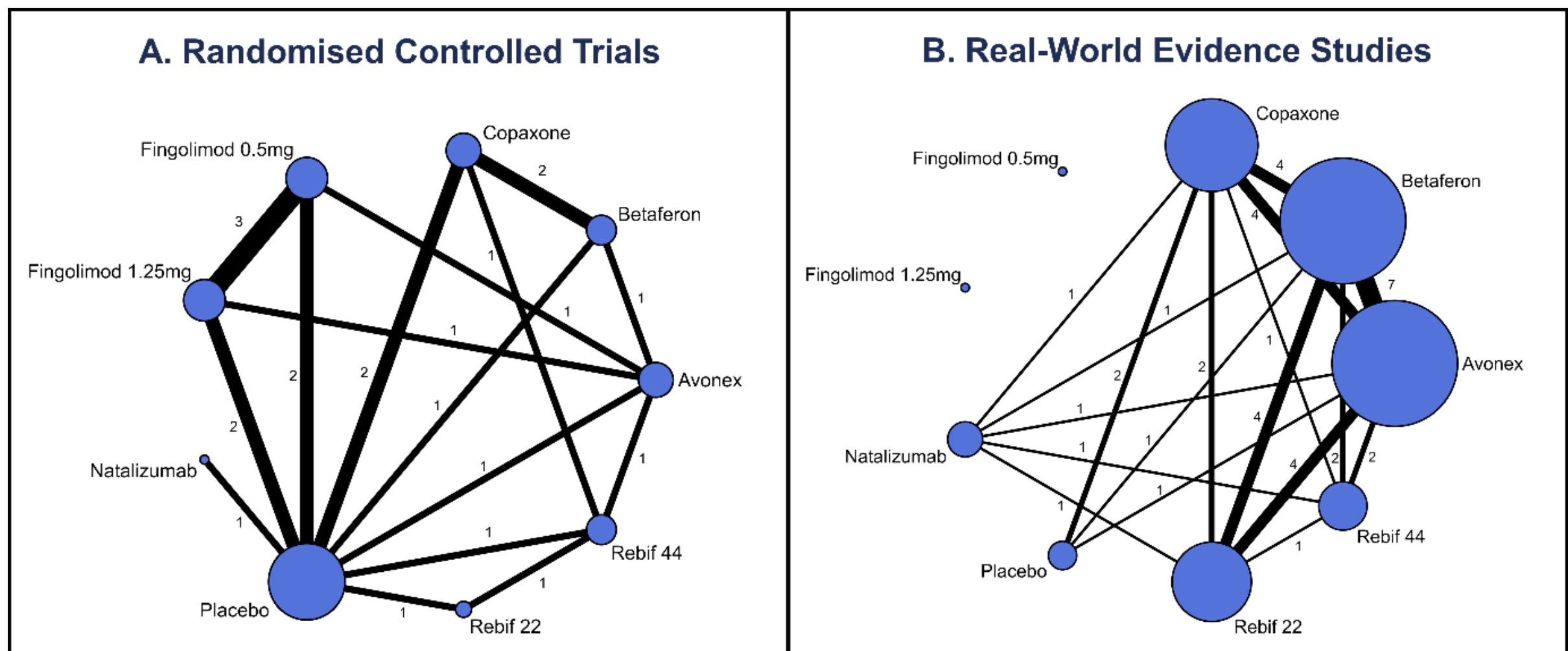

**Figure 1 - Network diagram including A) randomised controlled trials (RCT) and B) real-world evidence (RWE) studies for the treatment of relapsing remitting multiple sclerosis**

*Nodes (circles) in the diagram represent treatments included in the network meta-analysis, with node sizes being proportional to the number of subjects in each treatment arm. Edges (lines between nodes) represent the direct comparisons available between treatments with thickness of edges being proportional to the number of direct comparisons available. Numbers along edges represent the number of studies directly comparing treatments.*

### Naive pooling using standard NMA

Table 1 shows the annualised relapse rate ratios (ARRRs) (95% credible intervals) for an NMA of RCTs only (lower triangle) and an NMA of both sources of evidence with no adjustments for study design (upper triangle). As seen in Table 1, the ARRRs comparing all treatments vs. placebo are less than one, indicating a relative reduction in ARRRs for all active treatments compared to placebo.

When NMA treatment effect estimates are based on both sources of evidence, the levels of uncertainty can increase. For example, when comparing the effectiveness of fingolimod 0.5mg with Avonex the 95% credible interval of the ARRR increased from (1.64 to 2.38) when using only RCT data, to (1.44 to 2.52) when combined data from both sources of evidence were used. This is likely due to the increased between-study heterogeneity, when the two different sources of evidence were combined.

**Table 1: Matrix table of annualised relapse rate ratios (95% credible intervals) for network meta-analysis (NMA) using naïve pooling random-effects models***

| Treatment | Placebo | Natalizumab | Fingolimod 1.25 | Fingolimod 0.5 | Avonex | Rebif 22 | Rebif 44 | Copaxone | Betaferon |
|---|---|---|---|---|---|---|---|---|---|
| **Placebo** | | 0.41 (0.29, 0.57) | 0.46 (0.35, 0.59) | 0.42 (0.32, 0.53) | 0.78 (0.65, 0.94) | 0.77 (0.61, 0.95) | 0.75 (0.60, 0.93) | 0.60 (0.50, 0.71) | 0.70 (0.58, 0.84) |
| **Natalizumab** | 0.32 (0.26, 0.38) | | 1.16 (0.74, 1.68) | 1.05 (0.67, 1.52) | 1.99 (1.34, 2.74) | 1.95 (1.30, 2.72) | 1.90 (1.28, 2.64) | 1.53 (1.02, 2.11) | 1.78 (1.20, 2.46) |
| **Fingolimod 1.25** | 0.46 (0.40, 0.54) | 1.48 (1.15, 1.89) | | 0.91 (0.69, 1.17) | 1.73 (1.30, 2.27) | 1.70 (1.22, 2.32) | 1.66 (1.20, 2.26) | 1.33 (0.97, 1.77) | 1.56 (1.14, 2.07) |
| **Fingolimod 0.5** | 0.42 (0.36, 0.49) | 1.36 (1.04, 1.37) | 0.92 (0.77, 1.08) | | 1.92 (1.44, 2.52) | 1.88 (1.35, 2.57) | 1.84 (1.33, 2.51) | 1.48 (1.08, 1.96) | 1.72 (1.27, 2.29) |
| **Avonex** | 0.83 (0.72, 0.96) | 2.67 (2.61, 3.38) | 1.81 (1.50, 2.16) | 1.98 (1.64, 2.38) | | 0.98 (0.80, 1.20) | 0.96 (0.78, 1.18) | 0.77 (0.63, 0.93) | 0.90 (0.76, 1.05) |
| **Rebif 22** | 0.72 (0.60, 0.86) | 2.32 (1.75, 2.99) | 1.57 (1.24, 1.97) | 1.72 (1.35, 2.16) | 0.87 (0.70, 1.08) | | 0.99 (0.78, 1.23) | 0.79 (0.62, 0.98) | 0.92 (0.74, 1.12) |
| **Rebif 44** | 0.68 (0.59, 0.78) | 2.18 (1.69, 2.75) | 1.48 (1.21, 1.80) | 1.62 (1.32, 1.97) | 0.82 (0.69, 0.96) | 0.95 (0.78, 1.14) | | 0.81 (0.64, 0.99) | 0.94 (0.75, 1.15) |
| **Copaxone** | 0.65 (0.57, 0.75) | 2.09 (1.62, 2.65) | 1.42 (1.16, 1.73) | 1.56 (1.27, 1.90) | 0.79 (0.66, 0.94) | 0.91 (0.73, 1.12) | 0.96 (0.82, 1.13) | | 1.17 (0.98, 1.41) |
| **Betaferon** | 0.67 (0.57, 0.77) | 2.15 (1.65, 2.71) | 1.45 (1.18, 1.77) | 1.59 (1.29, 1.96) | 0.81 (0.67, 0.95) | 0.93 (0.74, 1.16) | 0.99 (0.82, 1.17) | 1.03 (0.89, 1.17) | |

* lower triangle consists of results from NMA of randomised controlled trials (RCTs) only and upper triangle consists of results from naïve-pooling NMA of RCTs and real-world evidence (RWE)

Note: For the NMA of RCTs (lower triangle), ARRRs are reported as rows vs columns (i.e., Natalizumab vs placebo ARRR 0.32 (0.26, 0.38). For the NMA of RCTs and RWEs (upper triangle), ARRRs are reported as columns vs rows (i.e., Natalizumab vs placebo ARRR 0.41 (0.29, 0.57).

## Power prior

The impact of the 'power transform prior' approach on the estimates of ARRRs (of each treatment compared to placebo) obtained from an NMA including both RCTs and RWE can be seen in Figure 2. The ARRRs of each active treatment compared to placebo are shown for a range of values of the down-weighting factor (alpha) between zero (maximum down-weighting, i.e. RWE not included) and one (RWE considered at 'face-value') (Additional file 4). It can be seen that for most of the active treatments there is relatively little impact of assigning increasing weight to the RWE in terms of the point estimates for the ARRRs. However, the impact on uncertainty around these estimates was noticeable. For example,  considering fingolimod 0.5mg compared to placebo (Figure 2) for alpha value of 0.001, ARRR (95% credible interval) estimated was 0.42 (0.36, 0.50) while an alpha value of 1.0 resulted in ARRR of 0.41 (0.32, 0.53). Whilst the point estimate remains fairly stable, the 95% credible interval widens as more weight is given to the RWE.  This may seem counter-intuitive, as more evidence is being included in the analysis, and therefore uncertainty levels would be expected to decrease. However, in this random-effects NMA, the between-study heterogeneity increased when including RWE, reflecting the differences observed between RCTs and RWE studies. This is represented by an increased between-study variance and in turn increased uncertainty in specific treatment effect estimates (see the last column of the Table in Additional file 4). However, because this applies consistently across all treatments the net impact, in terms of treatment rankings, is minimal as can be seen in Figure 3.

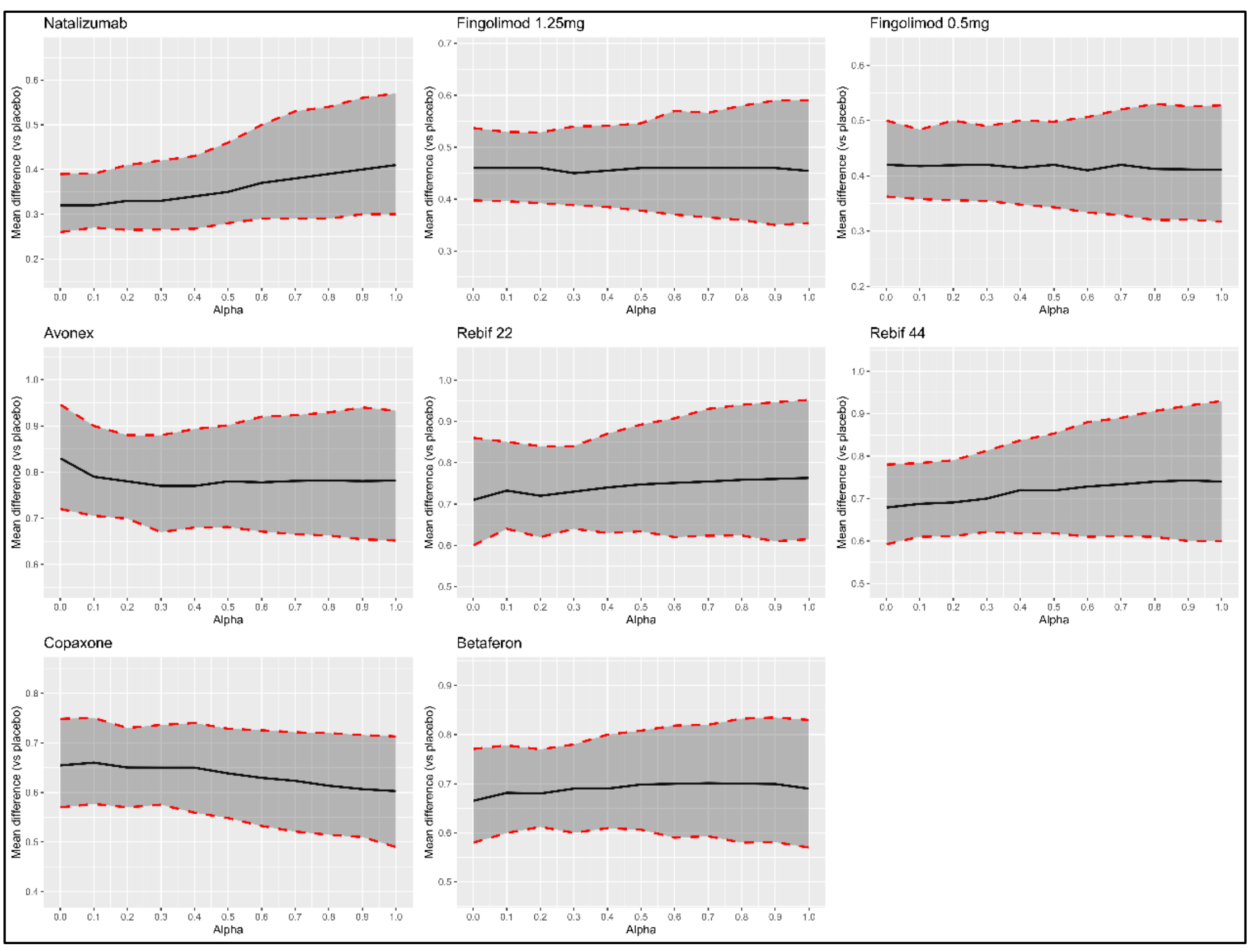

**Figure 2 - Annualised relapse rate ratios with 95% credible intervals for all active treatments compared to placebo for values of the down-weighting factor (alpha) using the power prior model**

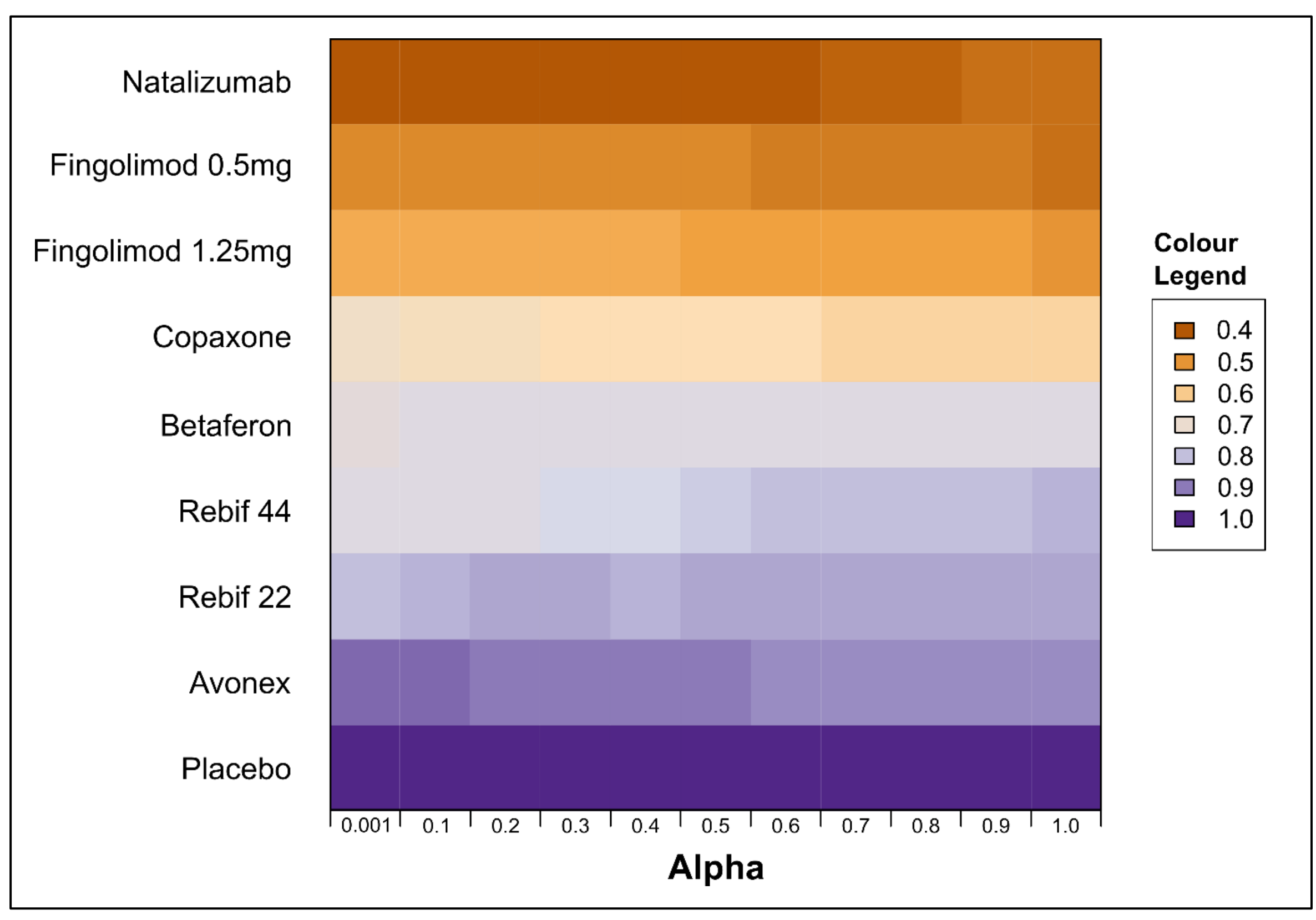

**Figure 3 - Heat map displaying rankings for each treatment (based on absolute annualised relapse rates) for values of the down-weighting factor (alpha) using power prior model**

*Orange represents highest ranking and purple represents lowest ranking.*

## Hierarchical model and hierarchical power prior model

Table 3 shows the results of adopting a hierarchical NMA which includes an additional level of hierarchy corresponding to the study design. Although the point estimates from the hierarchical model are in a broad agreement with the results presented above using a simpler 'power transform approach', it can be seen that the levels of uncertainty (in terms of the width of the credible intervals) are generally greater. For example, in comparison to placebo, natalizumab had an ARRR of 0.41 (0.30, 0.57) when using a power prior approach when alpha is 1 (Figure 2), while an ARRR of 0.40 (0.26, 0.70) in the hierarchical model (upper triangle in Table 3). This is due to the fact that the hierarchical model explicitly takes into account the differences between study designs, thus allowing for additional variability across studies. Extending the Hierarchical model to include 'power transform prior' approach ARRR effect estimates for a range of alpha values are included in Additional file 5. These differences in credible intervals were further observed in comparison to the power prior approach estimates. However, including RWE using the hierarchical model did not have any impact on the estimate of effectiveness for fingolimod (0.5mg and 1.25mg), which was due to the lack of RWE for this treatment and the nature of the model allowing for additional variability.

**Table 2: Matrix table of annualised relapse rate ratios (95% credible intervals) for network meta-analysis (NMA) using hierarchical models including randomised controlled trials and real-world evidence***

| Treatment | Placebo | Natalizumab | Fingolimod 1.25 | Fingolimod 0.5 | Avonex | Rebif 22 | Rebif 44 | Copaxone | Betaferon |
|---|---|---|---|---|---|---|---|---|---|
| **Placebo** | | 0.40 (0.26, 0.70) | 0.46 (0.40, 0.54) | 0.42 (0.36, 0.49) | 0.83 (0.55, 1.26) | 0.78 (0.51, 1.21) | 0.79 (0.53, 1.27) | 0.62 (0.40, 0.92) | 0.72 (0.48, 1.13) |
| **Natalizumab** | 0.35 (0.14, 0.74) | | 1.22 (0.65, 1.80) | 1.12 (0.60, 1.65) | 2.18 (1.04, 3.51) | 2.04 (0.99, 3.36) | 2.06 (1.04, 3.50) | 1.62 (0.75, 2.60) | 1.90 (0.93, 3.14) |
| **Fingolimod 1.25** | 0.46 (0.40, 0.54) | 1.76 (0.61, 3.40) | | 0.92 (0.76, 1.09) | 1.71 (1.16, 2.79) | 1.61 (1.09, 2.66) | 1.62 (1.12, 2.80) | 1.28 (0.84, 2.02) | 1.50 (1.02, 2.51) |
| **Fingolimod 0.5** | 0.42 (0.36, 0.49) | 1.60 (0.57, 3.13) | 0.92 (0.77, 1.07) | | 1.87 (1.27, 3.04) | 1.76 (1.18, 2.91) | 1.77 (1.22, 3.05) | 1.39 (0.93, 2.21) | 1.64 (1.11, 2.73) |
| **Avonex** | 0.88 (0.44, 1.60) | 3.39 (0.99, 7.53) | 1.70 (0.96, 3.48) | 1.86 (1.05, 3.79) | | 0.98 (0.54, 1.67) | 0.99 (0.56, 1.76) | 0.71 (0.41, 1.29) | 0.84 (0.51, 1.56) |
| **Rebif 22** | 0.79 (0.39, 1.56) | 3.22 (0.92, 6.58) | 1.50 (0.83, 3.43) | 1.64 (0.91, 3.73) | 1.00 (0.40, 2.11) | | 1.06 (0.59, 1.85) | 0.76 (0.43, 1.36) | 0.89 (0.53, 1.65) |
| **Rebif 44** | 0.76 (0.40, 1.50) | 3.17 (0.86, 6.86) | 1.46 (0.86, 3.28) | 1.60 (0.93, 3.57) | 0.97 (0.38, 2.16) | 1.09 (0.41, 2.38) | | 0.75 (0.41, 1.31) | 0.88 (0.50, 1.59) |
| **Copaxone** | 0.68 (0.34, 1.22) | 2.98 (0.73, 5.65) | 1.32 (0.72, 2.68) | 1.45 (0.79, 2.94) | 0.69 (0.31, 1.82) | 0.76 (0.33, 2.01) | 0.79 (0.33, 2.10) | | 1.23 (0.70, 2.16) |
| **Betaferon** | 0.74 (0.39, 1.41) | 2.91 (0.87, 6.49) | 1.43 (0.84, 3.08) | 1.56 (0.91, 3.39) | 0.75 (0.38, 1.99) | 0.84 (0.40, 2.25) | 0.86 (0.40, 2.30) | 1.23 (0.49, 2.74) | |

* lower triangle consists of results from NMA of randomised controlled trials (RCTs) only and upper triangle consists of results from hierarchical NMA of RCTs and real-world evidence (RWE)

Note: For the hierarchical NMA of RCTs (lower triangle), ARRRs are reported as rows vs columns (i.e., Natalizumab vs placebo ARRR 0.35 (0.14, 0.74). For the hierarchical NMA of RCTs and RWEs (upper triangle), ARRRs are reported as columns vs rows (i.e., Natalizumab vs placebo ARRR 0.40 (0.26, 0.70).

## Discussion

As previous research has suggested, there are differences between RCTs and RWE studies (Ioannidis et al., 2001). However, the results from this study did not show that including the RWE simply over- or underestimated the treatment effect for each treatment, but rather that there was both over- and underestimation for different treatments, supporting previous findings (Schmitz et al., 2013).

This study has further extended the methods introduced by Schmitz et al. (2013), by adapting them to model count data with the Poisson likelihood as well as extending the hierarchical model to down-weight the observational studies using a modified power prior approach. Both the hierarchical model and the modified hierarchical model are useful as they account for the heterogeneity between study designs and potential bias in RWE studies in the case of the latter (Turner et al., 2012). However, the results of these analyses did not differ significantly from the naïve pooling results or basic power transform prior results for this illustrative example. They also produced wider credible intervals due to the increased between-study design heterogeneity when including RWE. Whilst the hierarchical models may be considered more appropriate (in that they account for differences in sources of heterogeneity) care needs to be taken, and it is advised to compare the results with those from the naïve pooling in a sensitivity analysis to assess how results differ in practice.

In our illustrative example the inclusion of RWE increased the overall level of uncertainty in the treatment effects, supporting previous findings (Schmitz et al., 2013). For example, when looking at the effectiveness of fingolimod 0.5mg in the general population, greater heterogeneity was observed across different RWE studies, resulting in additional uncertainty around the effectiveness of fingolimod in the combined analysis in comparison to the effectiveness based on the carefully selected population in RCTs. The inclusion of RWE may increase the overall level of heterogeneity, and thus the uncertainty in estimated treatment effects – as was the case here. Thus, further evaluation of such methods in other settings, including the use of simulation studies, is warranted, and extension of the hierarchical modelling approach to allow for *different* types of RWE, either by inclusion of study-level covariates or by adding an extra level into the hierarchy, may ameliorate any potential increase in uncertainty regarding the treatment effects due to increased heterogeneity due to a broader evidence base (Saramago et al., 2012, Simmonds et al., 2005, Thom et al., 2015).

Implications for decision makers are that the methods can allow them to undertake assessments on a larger evidence base, and which includes a wider range of patient demographics and clinical characteristics. The inclusion of RWE in appraising health technologies can provide a larger (and possibly more representative) evidence base for decision-making; however, HTA analysts and decision-makers will need to consider on case-by-case basis whether or not the available RWE is sufficiently credible, whether this type of analysis is acceptable, and how the results should be interpreted and ultimately used.

### Limitations

There are a number of limitations of this study that need to be recognised. First, the sample sizes of RWE studies were smaller compared to the larger RCTs available for RRMS, which may have had an impact on the uncertainty of effect estimates when weighting studies. Second, this study has only utilised one illustrative example and results may differ in other clinical area.  While this may be the case, it remains of importance to compare the analysis of combined RCT and RWE data to the traditional NMA of RCT data alone to investigate the degree of effectiveness vs. efficacy gap. Thirdly, a Poisson likelihood was used to analyse this data. It is possible that the increased uncertainty could be reduced by utilising a negative binomial likelihood which can account for potential over dispersion when modelling count data. Fourth, meta-regression was not considered in this study. While meta-regression may explain some of the between-study heterogeneity, it may be limited both by the covariate information available and/or the number of studies in the NMA. Fifth, the NMAs in this particular example included aggregate level data only. Access to individual patient data from RWE would allow for adjustment of the results for potential allocation bias, potentially reducing the between-study heterogeneity and consequently the uncertainty around the pooled effectiveness estimates. However, obtaining IPD from observational studies can often be difficult due to the regulations around sharing such data. Further research would be needed to assess the impact of utilising IPD from observational studies. Finally, extraction of count data analysed with exact Poisson likelihood was considered more appropriate than, for example, extracting data on adjusted ARRRs (with modelling based on the normal approximation). However, this has its limitations as this prevents adjustment of treatment effects for confounding factors, which would only be possible with data at the IPD level.

## Conclusions

While the 'power transform prior' NMA as well as hierarchical NMA models had little impact on ARRR effect estimates, the degree of inclusion of RWE in the NMAs impacted the level of uncertainty around these effect estimates, likely as a result of increased between-study heterogeneity. The hierarchical NMA models provided another level of uncertainty, accounting to the differing study types (i.e. RCTs and RWE). Therefore, a comprehensive simulation study is required to investigate the ability of these models to correctly estimate treatment effects whilst also accounting for biases introduced by using RWE in different scenarios.

RWE can provide valuable data for HTA decision-making and in this paper we have illustrated a number of formal approaches for incorporating such data in evidence synthesis. Further, RWE can provide additional information, particularly in the case of rare diseases where clinical trial data are limited. Inclusion of RWE in meta-analysis can also be useful in clinical development planning as in Martina et al. (2018), who showed that inclusion of non-randomised data in meta-analysis can help inform the design of a future trial and potentially reduce the number of patients required as part of a drug development programme (Martina et al., 2018). However, the added value of RWE should be considered on a case-by-case basis.

# Competing interests

SB has served as a paid consultant, providing methodological advice, to NICE and Roche, and has received research funding from European Federation of Pharmaceutical Industries & Associations (EFPIA) and Johnson & Johnson.

KRA has served as a paid consultant, providing methodological advice, to; Abbvie, Amaris, Allergan, Astellas, AstraZeneca, Boehringer Ingelheim, Bristol-Meyers Squibb, Creativ-Ceutical, GSK, ICON/Oxford Outcomes, Ipsen, Janssen, Eli Lilly, Merck, NICE, Novartis, NovoNordisk, Pfizer, PRMA, Roche and Takeda, and has received research funding from Association of the British Pharmaceutical Industry (ABPI), European Federation of Pharmaceutical Industries & Associations (EFPIA), Pfizer, Sano_ and Swiss Precision Diagnostics. He is a Partner and Director of Visible Analytics Limited, a healthcare consultancy company.

All other authors declare that they have no competing interests

# Funding

The work leading to these results has received support from the Innovative Medicines Initiative Joint Undertaking under grant agreement n° [115546], resources of which are composed of financial contribution from the European Union's Seventh Framework Programme (FP7/2007- 2013) and EFPIA companies' in kind contribution. KRA, SB and HH were partially supported by the UK Medical Research Council [grant no. MR/R025223/1]. SB was partially supported by the Medical Research Council (MRC) Methodology Research Programme [New Investigator Research Grant MR/L009854/1]. KRA was partially supported as a NIHR Senior Investigator Emeritus (NI-SI-0512-10159). DAJ was partly supported by the National Institute for Health Research (NIHR) Greater Manchester Patient Safety Translational Research Centre (NIHR Greater Manchester PSTRC). The views expressed are those of the authors and not necessarily those of the Northern Health Science Alliance, the NHS, the NIHR or the Department of Health and Social Care.

# Authors' contribution

KRA, DJ and SB conceived the concept and design of this study. DJ and HH carried out the analyses. DJ drafted the first version of the manuscript. SB and HH critically reviewed and made substantial contributions to the subsequent versions of the manuscript. All authors contributed to the discussions throughout the research project and commented and approved subsequent manuscript drafts.

# Acknowledgments

The authors would like to thank Drs Melvin 'Skip' Olson and Alexandre Joyeux, both of Novartis &. IMI GetReal Work Package 1, for helpful discussions regarding the illustrative case study on relapsing remitting multiple sclerosis.

# Additional File 1

**Search terms used for the systematic review assessing the impact of treatments in relapsing remitting multiple sclerosis**

| Step | Terms Used |
|---|---|
| 1 | Multiple Sclerosis, Relapsing-Remitting/ or multiple sclerosis.sh,tw. |
| 2 | (relapsing remitting adj2 multiple sclerosis).ti,ab,sh,hw,ot. |
| 3 | (remitting relapsing adj2 multiple sclerosis).ti,ab,sh,hw,ot. |
| 4 | ((relapsing remitting adj2 ms) or (remitting relapsing adj2 ms)).ti,ab,sh,hw,ot. |
| 5 | ((exacerbat* or disseminated or insular or secondary progressive or primary progressive or progressive relapsing) adj2 (sclerosis or ms)).ti,ab,sh,hw,ot. |
| 6 | (rrms or encephalomyelitis disseminat*).ti,ab,sh,hw,ot. |
| 7 | or/1-6 |
| 8 | (teriflunomide or A 1726 or A 77 1726 or A 771726 or HMR 1726 or HMR1726 or aubagio or 163451-81-8).ti,ab,rn,nm,sh,hw,ot. |
| 9 | (fingolimod or FTY-720 or FTY270 or Gilenya or Gilenia or 162359-55-9).ti,ab,rn,nm,sh,hw,ot. |
| 10 | (natalizumab or tysabri or antegren or 189261-10-7).ti,ab,rn,nm,sh,hw,ot. |
| 11 | or/8-10 |
| 12 | Interferon-beta/ |
| 13 | (interferon-beta-1 or interferon-1a or interferon-1b or interferon beta or beta Interferon or Interferon beta 1 or 220581-49-7 or 145155-23-3).ti,ab,rn,nm,sh,hw,ot. |
| 14 | (avonex or rebif or betaferon or betaseron or BAY 86-5046 or BAY86-5046 or extavia).ti,ab,rn,nm,sh,hw,ot. |
| 15 | (glatiramer acetate or copaxone or 147245-92-9).ti,ab,rn,nm,sh,hw,ot. |
| 16 | (fampridine or fampyra or 504-24-5).ti,ab,rn,nm,sh,hw,ot. |
| 17 | or/12-16 |
| 18 | exp epidemiologic studies/ or (epidemiolog* adj (study or studies)).tw. |
| 19 | Observational Study.pt. or ((observational or cohort or case-control or case control or cross sectional or cross-sectional) adj (study or studies)).tw. |
| 20 | ((follow up or follow-up) adj2 (study or studies)).tw. |
| 21 | (longitudinal or retrospective or prospective).tw. |
| 22 | or/18-21 |
| 23 | Comparative study.pt. or (comparative adj (study or studies)).tw. |
| 24 | Multicenter Study.pt. |
| 25 | Pragmatic Clinical Trial.pt. or (pragmatic$ adj3 (study or studies or trial$)).tw. |
| 26 | Clinical Trial, Phase IV.pt. or (phase adj ((relapsing remitting adj2 ms) or (remitting relapsing adj2 ms) or IV or four) adj (study or studies or trial$)).tw. |
| 27 | exp Questionnaires/ |
| 28 | Health Care Surveys/ |
| 29 | Health Surveys/ |
| 30 | Registries/ or registr*.tw. |
| 31 | Records as Topic/ or (((hospital or medical) adj records) or (chart$ adj3 review$)).tw. |
| 32 | (database adj5 (study or studies or report* or research or activit*)).tw. |
| 33 | ((monitoring or surveillance or extension) adj3 (data or study or studies or report* or trial*)).tw. |
| 34 | or/23-33 |
| 35 | Meta-Analysis.pt. |
| 36 | meta-analysis/ or systematic review/ or meta-analysis as topic/ or "meta analysis (topic)"/ or "systematic review (topic)"/ or exp technology assessment, biomedical/ |
| 37 | (met analy* or metanaly* or technology assessment* or HTA or HTAs or technology overview*).ti,ab. |
| 38 | (meta-analy* or metaanaly* or systematic review* or biomedical technology assessment* or bio-medical technology assessment*).mp,hw. |
| 39 | or/35-38 |
| 40 | ((real world adj (data or evidence)) or relative effectiveness).tw. |
| 41* | 7 and (22 or 34 or 39 or 40) and 8 - TERIFLUNOMIDE |
| 42* | 7 and (22 or 34 or 39 or 40) and 9 - FINGOLIMOD |
| 43** | 7 and (22 or 34 or 39 or 40) and 10 - NATALIZUMAB |
| 44* | 7 and (22 or 34 or 39 or 40) and 12 and 13 and 14 - INTERFERON BETAS |
| 45* | 7 and (22 or 34 or 39 or 40) and 15 - GLATIRAMER ACETATE |
| 46* | 7 and (22 or 34 or 39 or 40) and 16 - FAMPRIDINE |

# Additional File 2

**Reference list of randomised controlled trials and real-world studies include in the network meta-analysis assessing the impact of treatments in relapsing remitting multiple sclerosis**

# Additional File 3

**Number of subjects, number of relapses and exposure time (person-years) extracted and analysed from randomised controlled trials and real-world studies assessing the impact of treatments in relapse remitting multiple sclerosis**

| First author | Treatment | No of subjects | No of relapses | Exposure time (person-years) |
|---|---|---|---|---|
| | | **Randomised controlled trials** | | |
| **Polman et al. (2006)** | Placebo | 315 | 433 | 593 |
| | Natalizumab | 627 | 276 | 1202 |
| **Kappos et al. (2010)** | Placebo | 418 | 300 | 750 |
| | Fingolimod (1.25mg) | 429 | 122 | 762 |
| | Fingolimod (0.5mg) | 425 | 143 | 794 |
| **Cohen et al. (2010)** | Avonex | 431 | 135 | 409 |
| | Fingolimod (1.25mg) | 420 | 79 | 395 |
| | Fingolimod (0.5mg) | 429 | 66 | 414 |
| **Freedoms (2010)** | Placebo | 355 | 0 | 710 |
| | Fingolimod (0.5mg) | 358 | 150 | 716 |
| | Fingolimod (1.25mg) | 370 | 0 | 740 |
| **PRISMS Group (1998)** | Placebo | 187 | 479 | 364 |
| | Rebif 22 | 189 | 344 | 366 |
| | Rebif 44 | 184 | 318 | 363 |
| **Johnson et al. (1995)** | Placebo | 126 | 210 | 250 |
| | Copaxone | 125 | 161 | 273 |
| **Comi et al. (2001)** | Placebo | 120 | 91 | 75 |
| | Copaxone | 119 | 61 | 75 |
| **Jacobs et al. (1996)** | Placebo | 143 | 225 | 274 |
| | Avonex | 158 | 196 | 293 |
| **Durelli et al. (2002)** | Avonex | 92 | 126 | 180 |
| | Betaferon | 96 | 95 | 190 |
| **Panitch et al. (2002)** | Avonex | 338 | 216 | 304 |
| | Rebif 44 | 339 | 183 | 305 |
| **IFNB Group (1993)** | Placebo | 123 | 266 | 209 |
| | Betaferon | 124 | 173 | 206 |
| **O'Connor et al. (2009)** | Copaxone | 448 | 383 | 1126 |
| | Betaferon | 897 | 828 | 2299 |
| **O'Connor et al. (2009)** | Copaxone | 39 | 23 | 70 |
| | Betaferon | 36 | 25 | 68 |
| **Mikol et al. (2008)** | Copaxone | 386 | 200 | 688 |
| | Rebif 44 | 387 | 207 | 689 |
| | | **Real-world evidence** | | |
| **Lanzillo (2010/2011)** | Natalizumab | 42 | 10 | 42 |
| | Rebif 44 | 42 | 23 | 42 |
| **Limmroth (2007) (QUASIMS)** | Avonex | 1094 | 1116 | 2188 |
| | Betaferon | 1034 | 1075 | 2068 |
| | Rebif 22 | 555 | 588 | 1110 |
| | Rebif 44 | 185 | 233 | 370 |
| **Halpern 2011** | Natalizumab | 288 | 21 | 72 |
| | Avonex | 151 | 7 | 38 |
| | Rebif 22 | 329 | 22 | 82 |
| | Betaferon | 144 | 11 | 36 |
| | Copaxone | 469 | 25 | 117 |
| **Patti (2006)** | Betaferon | 114 | 137 | 570 |
| | Avonex | 37 | 50 | 185 |
| | Rebif 22 | 17 | 35 | 85 |
| **Rio (2005)** | Placebo | 107 | 288 | 356 |
| | Copaxone | 101 | 204 | 334 |
| **Haas and Firzlaff (2005)** | Avonex | 79 | 109 | 158 |
| | Betaferon | 77 | 123 | 154 |
| | Copaxone | 79 | 56 | 158 |
| | Rebif 22 | 48 | 59 | 96 |
| **Khan et al. (2001)** | Placebo | 15 | 23 | 23 |
| | Avonex | 34 | 41 | 51 |
| | Betaferon | 34 | 28 | 51 |
| | Copaxone | 39 | 29 | 59 |
| **Trojano et al. (2003)** | Betaferon | 209 | 136 | 418 |
| | Avonex | 169 | 120 | 338 |
| **Carra et al. (2003)** | Avonex | 26 | 14 | 35 |
| | Rebif 44 | 20 | 12 | 27 |
| | Betaferon | 20 | 11 | 27 |
| | Copaxone | 30 | 8 | 40 |

# Additional File 4

**Annualised relapse rate ratios (95% credible intervals) of each active treatment compared to placebo for values alpha using the power prior model with between study heterogeneity standard deviation estimates**

| Alpha | **Natalizumab** | **Fingolimod 1.25mg** | **Fingolimod 0.5mg** | **Avonex** | **Rebif 22** | **Rebif 44** | **Copaxone** | **Betaferon** | **Between study SD** |
|---|---|---|---|---|---|---|---|---|---|
| 0.001 | 0.32 (0.26, 0.39) | 0.46 (0.40, 0.54) | 0.42 (0.36, 0.50) | 0.83 (0.72, 0.95) | 0.71 (0.60, 0.86) | 0.68 (0.59, 0.78) | 0.65 (0.57, 0.75) | 0.67 (0.58, 0.77) | 0.055 |
| 0.1 | 0.32 (0.27, 0.39) | 0.46 (0.40, 0.53) | 0.42 (0.36, 0.48) | 0.79 (0.71, 0.90) | 0.73 (0.64, 0.85) | 0.69 (0.61, 0.78) | 0.66 (0.58, 0.75) | 0.68 (0.60, 0.78) | 0.045 |
| 0.2 | 0.33 (0.27, 0.41) | 0.46 (0.39, 0.53) | 0.42 (0.36, 0.50) | 0.78 (0.70, 0.88) | 0.72 (0.62, 0.84) | 0.69 (0.61, 0.79) | 0.65 (0.57, 0.73) | 0.68 (0.61, 0.77) | 0.047 |
| 0.3 | 0.33 (0.27, 0.42) | 0.45 (0.39, 0.54) | 0.42 (0.35, 0.49) | 0.77 (0.67, 0.88) | 0.73 (0.64, 0.84) | 0.70 (0.62, 0.81) | 0.65 (0.58, 0.74) | 0.69 (0.60, 0.78) | 0.057 |
| 0.4 | 0.34 (0.27, 0.43) | 0.45 (0.38, 0.54) | 0.41 (0.35, 0.50) | 0.77 (0.68, 0.89) | 0.74 (0.63, 0.87) | 0.72 (0.62, 0.84) | 0.65 (0.56, 0.74) | 0.69 (0.61, 0.80) | 0.085 |
| 0.5 | 0.35 (0.28, 0.46) | 0.46 (0.38, 0.55) | 0.42 (0.34, 0.50) | 0.78 (0.68, 0.90) | 0.75 (0.63, 0.89) | 0.72 (0.62, 0.85) | 0.64 (0.55, 0.73) | 0.70 (0.61, 0.81) | 0.100 |
| 0.6 | 0.37 (0.29, 0.50) | 0.46 (0.37, 0.57) | 0.41 (0.33, 0.51) | 0.78 (0.67, 0.92) | 0.75 (0.62, 0.91) | 0.73 (0.61, 0.88) | 0.63 (0.53, 0.73) | 0.70 (0.59, 0.82) | 0.131 |
| 0.7 | 0.38 (0.29, 0.53) | 0.46 (0.36, 0.57) | 0.42 (0.33, 0.52) | 0.78 (0.67, 0.92) | 0.75 (0.62, 0.93) | 0.73 (0.61, 0.89) | 0.62 (0.52, 0.72) | 0.70 (0.59, 0.82) | 0.144 |
| 0.8 | 0.39 (0.29, 0.54) | 0.46 (0.36, 0.58) | 0.41 (0.32, 0.53) | 0.78 (0.66, 0.93) | 0.76 (0.62, 0.94) | 0.74 (0.61, 0.91) | 0.61 (0.51, 0.72) | 0.70 (0.58, 0.83) | 0.162 |
| 0.9 | 0.40 (0.30, 0.56) | 0.46 (0.35, 0.59) | 0.41 (0.32, 0.53) | 0.78 (0.65, 0.94) | 0.76 (0.61, 0.95) | 0.74 (0.60, 0.92) | 0.61 (0.51, 0.72) | 0.70 (0.58, 0.83) | 0.173 |
| 1.0 | 0.41 (0.30, 0.57) | 0.45 (0.35, 0.59) | 0.41 (0.32, 0.53) | 0.78 (0.65, 0.93) | 0.76 (0.61, 0.95) | 0.74 (0.60, 0.93) | 0.60 (0.49, 0.71) | 0.69 (0.57, 0.83) | 0.182 |

# Additional File 5

**Annualised relapse rate ratios (95% credible intervals) of each active treatment compared to placebo for values of the down-weighting factor (alpha) between zero (total down-weighting, i.e. RWE not included) and one (RWE considered at 'face-value') using the hierarchical power prior model**

| Alpha | Natalizumab | Fingolimod 1.25mg | Fingolimod 0.5mg | Avonex | Rebif 22 | Rebif 44 | Copaxone | Betaferon |
|---|---|---|---|---|---|---|---|---|
| **0.001** | 0.35 (0.14, 0.74) | 0.46 (0.40, 0.54) | 0.42 (0.36, 0.49) | 0.88 (0.44, 1.60) | 0.79 (0.39, 1.56) | 0.76 (0.40, 1.50) | 0.68 (0.34, 1.22) | 0.74 (0.39, 1.41) |
| **0.1** | 0.33 (0.24, 0.50) | 0.46 (0.40, 0.53) | 0.42 (0.36, 0.49) | 0.81 (0.61, 1.07) | 0.75 (0.57, 1.04) | 0.73 (0.56, 1.05) | 0.65 (0.48, 0.84) | 0.70 (0.54, 0.97) |
| **0.2** | 0.35 (0.25, 0.54) | 0.46 (0.40, 0.53) | 0.42 (0.36, 0.49) | 0.80 (0.60, 1.09) | 0.75 (0.55, 1.02) | 0.74 (0.56, 1.07) | 0.63 (0.46, 0.84) | 0.71 (0.53, 0.98) |
| **0.3** | 0.35 (0.25, 0.56) | 0.46 (0.40, 0.53) | 0.42 (0.36, 0.49) | 0.81 (0.60, 1.10) | 0.75 (0.56, 1.04) | 0.74 (0.56, 1.07) | 0.63 (0.46, 0.83) | 0.71 (0.53, 0.97) |
| **0.4** | 0.37 (0.25, 0.61) | 0.46 (0.40, 0.54) | 0.42 (0.36, 0.49) | 0.82 (0.59, 1.18) | 0.77 (0.55, 1.12) | 0.77 (0.56, 1.17) | 0.63 (0.44, 0.87) | 0.72 (0.52, 1.05) |
| **0.5** | 0.37 (0.26, 0.59) | 0.46 (0.40, 0.54) | 0.42 (0.36, 0.49) | 0.82 (0.59, 1.17) | 0.77 (0.55, 1.11) | 0.77 (0.56, 1.17) | 0.63 (0.44, 0.87) | 0.72 (0.52, 1.03) |
| **0.6** | 0.40 (0.26, 0.73) | 0.46 (0.40, 0.54) | 0.42 (0.36, 0.49) | 0.85 (0.55, 1.40) | 0.80 (0.52, 1.32) | 0.81 (0.54, 1.37) | 0.64 (0.41, 1.00) | 0.75 (0.49, 1.24) |
| **0.7** | 0.40 (0.26, 0.73) | 0.46 (0.40, 0.54) | 0.42 (0.36, 0.49) | 0.84 (0.55, 1.36) | 0.79 (0.52, 1.29) | 0.81 (0.54, 1.37) | 0.63 (0.41, 0.98) | 0.74 (0.49, 1.22) |
| **0.8** | 0.39 (0.26, 0.68) | 0.46 (0.39, 0.54) | 0.42 (0.36, 0.49) | 0.83 (0.55, 1.26) | 0.77 (0.52, 1.19) | 0.78 (0.53, 1.26) | 0.62 (0.41, 0.91) | 0.72 (0.49, 1.11) |
| **0.9** | 0.40 (0.26, 0.68) | 0.46 (0.40, 0.54) | 0.42 (0.36, 0.49) | 0.82 (0.55, 1.25) | 0.77 (0.51, 1.21) | 0.79 (0.53, 1.27) | 0.62 (0.40, 0.92) | 0.72 (0.49, 1.11) |
| **1.0** | 0.40 (0.26, 0.70) | 0.46 (0.40, 0.54) | 0.42 (0.36, 0.49) | 0.83 (0.55, 1.26) | 0.78 (0.51, 1.21) | 0.79 (0.53, 1.27) | 0.62 (0.40, 0.92) | 0.72 (0.48, 1.13) |